\documentclass[twocolumn,prl,amsmath,amssymb,nopacs,superscriptaddress]{revtex4}
\usepackage{graphicx}
\usepackage{bm}
\usepackage{latexsym}
\usepackage{amstext,amssymb}
\usepackage{amsxtra,amsmath}
\usepackage{color}

\newcommand{\bs}{\boldsymbol}

\usepackage{times}

\newcommand{\ket}[1]{|#1\rangle}

\newcommand{\bibl}[1]{$^{#1}$}

\makeatletter
\makeatletter \renewcommand{\fnum@figure}
{\textbf{\small\figurename~\thefigure ${\bs |}$}}
\makeatother

\begin{document}

\title{ Realizing quantum Ising models in tunable two-dimensional arrays of single Rydberg atoms}

\author{Henning Labuhn}
\affiliation{ Laboratoire Charles Fabry, Institut d'Optique, CNRS, Univ Paris Sud 11, 2 avenue Augustin Fresnel, 91127 Palaiseau cedex, France}

\author{Daniel Barredo}
\affiliation{ Laboratoire Charles Fabry, Institut d'Optique, CNRS, Univ Paris Sud 11, 2 avenue Augustin Fresnel, 91127 Palaiseau cedex, France}

\author{Sylvain Ravets}
\affiliation{ Laboratoire Charles Fabry, Institut d'Optique, CNRS, Univ Paris Sud 11, 2 avenue Augustin Fresnel, 91127 Palaiseau cedex, France}

\author{Sylvain~de~L\'es\'eleuc}
\affiliation{ Laboratoire Charles Fabry, Institut d'Optique, CNRS, Univ Paris Sud 11, 2 avenue Augustin Fresnel, 91127 Palaiseau cedex, France}

\author{Tommaso Macr\`i}
\affiliation{Departamento de F\'isica Te\'orica e Experimental, Universidade Federal do Rio Grande do Norte, and International Institute of Physics, Natal-RN, Brazil. }

\author{Thierry Lahaye}
\affiliation{ Laboratoire Charles Fabry, Institut d'Optique, CNRS, Univ Paris Sud 11, 2 avenue Augustin Fresnel, 91127 Palaiseau cedex, France}

\author{Antoine Browaeys}
\affiliation{ Laboratoire Charles Fabry, Institut d'Optique, CNRS, Univ Paris Sud 11, 2 avenue Augustin Fresnel, 91127 Palaiseau cedex, France}

\maketitle

{\bf 
Spin models are the prime example of simplified many-body Hamiltonians used to model complex, real-world strongly correlated materials\bibl{1}. However, despite their simplified character, their dynamics often cannot be simulated exactly on classical computers as soon as the number of particles exceeds a few tens. For this reason, the quantum simulation\bibl{2} of spin Hamiltonians using the tools of atomic and molecular physics has become very active over the last years, using ultracold atoms\bibl{3} or molecules\bibl{4} in optical lattices, or trapped ions\bibl{5}. All of these approaches have their own assets, but also limitations. Here, we report on a novel platform for the study of spin systems, using individual atoms trapped in two-dimensional arrays of optical microtraps with arbitrary geometries, where filling fractions range from $60$ to $100\%$ with exact knowledge of the initial configuration. When excited to Rydberg $D$-states, the atoms undergo strong interactions whose anisotropic character opens exciting prospects for simulating exotic matter\bibl{6}. We illustrate the versatility of our system by studying the dynamics of an Ising-like spin-$1/2$ system in a transverse field with up to thirty spins, for a variety of geometries in one and two dimensions, and for a wide range of interaction strengths. For geometries where the anisotropy is expected to have small effects we find an excellent agreement with \emph{ab-initio} simulations of the spin-$1/2$ system, while for strongly anisotropic situations the multilevel structure of the $D$-states has a measurable influence\bibl{7,8}. Our findings establish arrays of single Rydberg atoms as a versatile platform for the study of quantum magnetism. 
}

Rydberg atoms have recently attracted a lot of interest for quantum information processing\bibl{9} and quantum simulation\bibl{10}. In this work, we use a system of individual Rydberg atoms to realize highly-tunable artificial quantum Ising magnets. By shining on the atoms lasers that are resonant with the transition between the ground state $\ket{g}$ and a chosen Rydberg state $\ket{r}$, we implement the Ising-like Hamiltonian
\begin{equation}
H=\sum_i\frac{\hbar\Omega}{2} \sigma_x^{i}+\sum_{i<j}V_{ij}n^{i}n^{j},
\label{eq:ising}
\end{equation}
which acts on the pseudo-spin states $\ket{\!\!\downarrow}_i$ and $\ket{\!\!\uparrow}_i$ corresponding to states $\ket{g}$ and $\ket{r}$ of atom $i$, respectively. Here, $\Omega$ is the Rabi frequency of the laser coupling, the $\sigma_\alpha^{i}$ ($\alpha=x,y,z$) are the Pauli matrices acting on atom $i$, and $n^{i}=(1+\sigma_z^{i})/2$ is the number of Rydberg excitations (0 or 1) on site $i$. The term $V_{ij}$ arises from the van der Waals interaction between atoms $i$ and $j$ when they are both in $\ket{r}$, and scales as $C_6(\theta)|{\boldsymbol r}_i-{\boldsymbol r}_j|^{-6}$ with the separation between the atoms. Moreover, for $\ket{r}=\ket{nD_{3/2},m_j=3/2}$, the interaction strength is anisotropic\bibl{7,11}, varying by $\sim 3$ when the angle $\theta$ between the interatomic axis and the quantization axis $\hat{z}$ changes from $0$ to $\pi/2$.

\begin{figure}
\centering
\includegraphics[width=8.5cm]{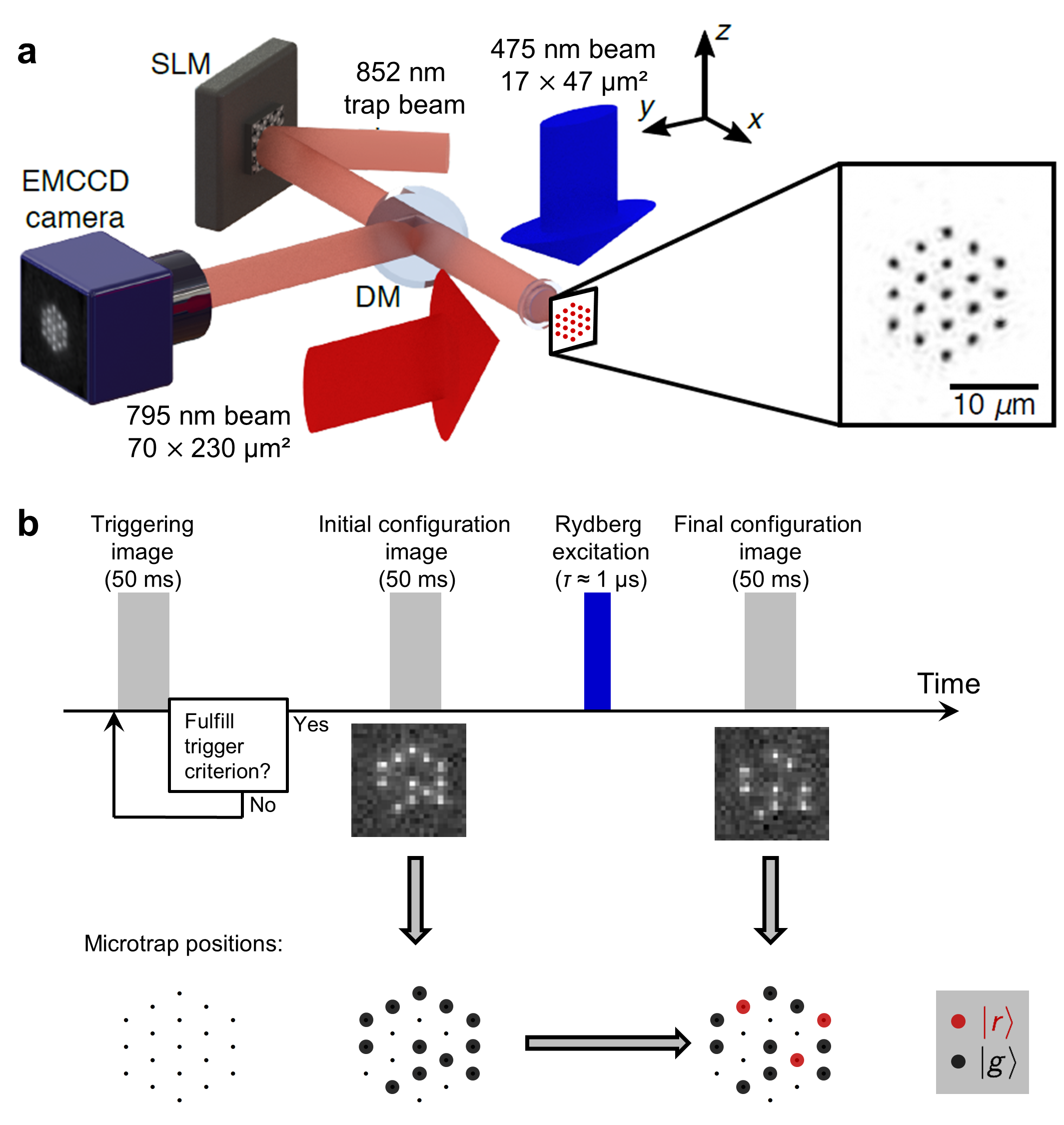}
\caption{\small {\bf Experimental platform.} {\bf a}: An array of microtraps is created by imprinting an appropriate phase on a dipole-trap beam. Site-resolved fluorescence of the atoms, at 780~nm, is imaged on a camera using a dichroic mirror (DM). Rydberg excitation beams at 795 and 475~nm are shone onto the atoms. The inset shows the measured light intensity for an array of $N_t=19$ traps. {\bf b}: Sketch of an experimental sequence. During loading, the camera images are analyzed continuously to extract the number of loaded traps. As soon as a triggering criterion is met, the loading is stopped and an image of the initial configuration is acquired. After Rydberg excitation, a final image is acquired, revealing the atoms excited to Rydberg states (red disks).  }
\label{fig:fig1}
\end{figure}

Our setup (Fig.~\ref{fig:fig1}a) has been described in refs.~12,13. We trap cold ($T\simeq 30\;\mu{\rm K}$) single $^{87}{\rm Rb}$ atoms in optical traps with a $1\;\mu{\rm m}$ waist. Using a spatial light modulator (SLM), we create arbitrary, two-dimensional arrays containing $1\leqslant N_{\rm t}\leqslant 50$ traps, separated by distances $a>3\;\mu{\rm m}$. The atomic fluorescence at 780~nm is imaged onto a camera. We observe, in the single-atom regime\bibl{12}, that the level of fluorescence for each trap alternates randomly between two levels, corresponding to the presence of 0 or 1 atom. The analysis of these $N_{\rm t}$ fluorescence traces allows us to record, with a time resolution of 50~ms, the current number $N$ of single atoms in the array. 

As illustrated in Figure~\ref{fig:fig1}b, as soon as $N$ exceeds a predefined threshold, we trigger the following experimental sequence. First, the loading of the array is stopped, and a fluorescence image is acquired to record the \emph{initial configuration} of the atoms, i.e. which traps are filled. After initializing all the atoms in $\ket{g}=\ket{5S_{1/2},F=2,m_F=2}$ by optical pumping, a two-photon Rydberg excitation pulse of duration $\tau$ is shone onto the atoms; the Rabi frequency ($\Omega\simeq2\pi\times 1\;{\rm MHz}$) is uniform to within 10\,\% over the array. We then acquire a new image, of the \emph{final configuration}. Atoms excited to $\ket{r}$ have quickly escaped the trapping region, and thus  we observe only the atoms that were in $\ket{g}$ after excitation. The atoms gone in between the initial and final images are thus assigned to Rydberg states (red dots in Figure~\ref{fig:fig1}b). This detection method has a high efficiency: it only gives a small number of `false positives', as an atom also has a probability $\varepsilon\simeq(3\pm1)\%$ to be lost, independently of its internal state (Methods).
 
\begin{figure}[t!]
\centering
\includegraphics[width=8.7cm]{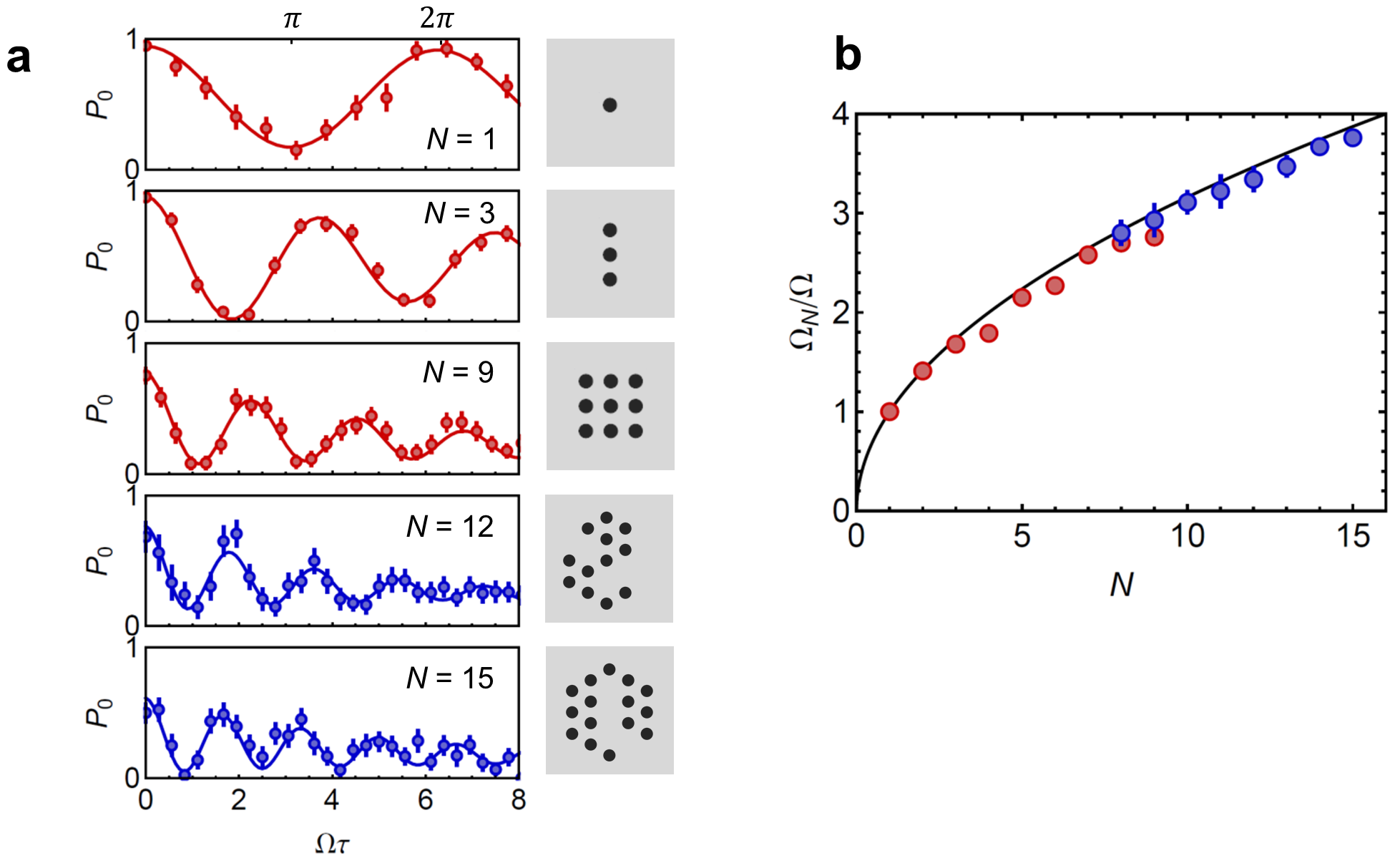}
\caption{\small {\bf Collective oscillations in the full Rydberg blockade regime.} {\bf a}: Probability $P_0$ for all $N$ atoms to be in $\ket{g}$ after an excitation pulse of area $\Omega \tau$. Red points: fully loaded arrays, $n=82$; blue points: partially loaded triangular arrays of $N_{\rm t}=19$ traps, $n=100$ (error bars show the quantum projection noise for $\sim 100$ repetitions of the experiment). Solid lines are fits by damped sines of frequency $\Omega_N$. {\bf b}: Collective oscillation frequency $\Omega_N/\Omega$ versus $N$ (error bars ---sometimes smaller than the symbol size--- are statistical). The solid line is the expected $\sqrt{N}$ enhancement. }
\label{fig:fig2}
\end{figure}

We first test our system in the conceptually simple situation of fully Rydberg-blockaded ensembles containing up to $N=15$ atoms. Figure~\ref{fig:fig2}a shows, for various arrays, the probability that all $N$ atoms are in $\ket{g}$ at the end of the sequence. We observe high-contrast coherent oscillations, with a frequency enhanced by a factor $\sqrt{N}$ with respect to the  single-atom case (Fig.~\ref{fig:fig2}b). This characteristic collective oscillation is the hallmark of Rydberg blockade\bibl{14-16}, where multiple excitations are inhibited within a blockaded volume (which, due to the anisotropy, is close to an ellipsoid, with a major radius $R_{\rm b}$ defined by $\hbar\Omega = |C_6(0)|/R_{\rm b}^6$, and a small `flattening' $3^{1/6}\simeq1.2$). This observation is a first step towards the creation of long-lived $\ket{W}$ states in the ground state\bibl{9}.

The fully blockaded regime remains easy to describe theoretically as the blockade naturally truncates the size of the Hilbert space. In contrast, a more challenging regime corresponds to the Rydberg blockade being effective only between nearest neighbors, such that for long enough excitation times, the number of excitations becomes $\sim N/2$. It is therefore desirable to be able to vary the ratio $\alpha=R_{\rm b}/a$ of the blockade radius to the distance $a$ between sites: for very small or large values of $\alpha$, the dynamics is simple and the system can easily be compared to numerics, while, for intermediate values of $\alpha$, the dynamics is challenging to calculate and experimental quantum simulation becomes relevant. Our setup is particularly adapted to this goal, as we can vary easily both $a$ (reconfiguring the SLM) and $R_{\rm b}$ (changing the principal quantum number $n$, we tune $C_6$ which scales approximately as $n^{11}$).

\begin{figure}
\centering
\includegraphics[width=85mm]{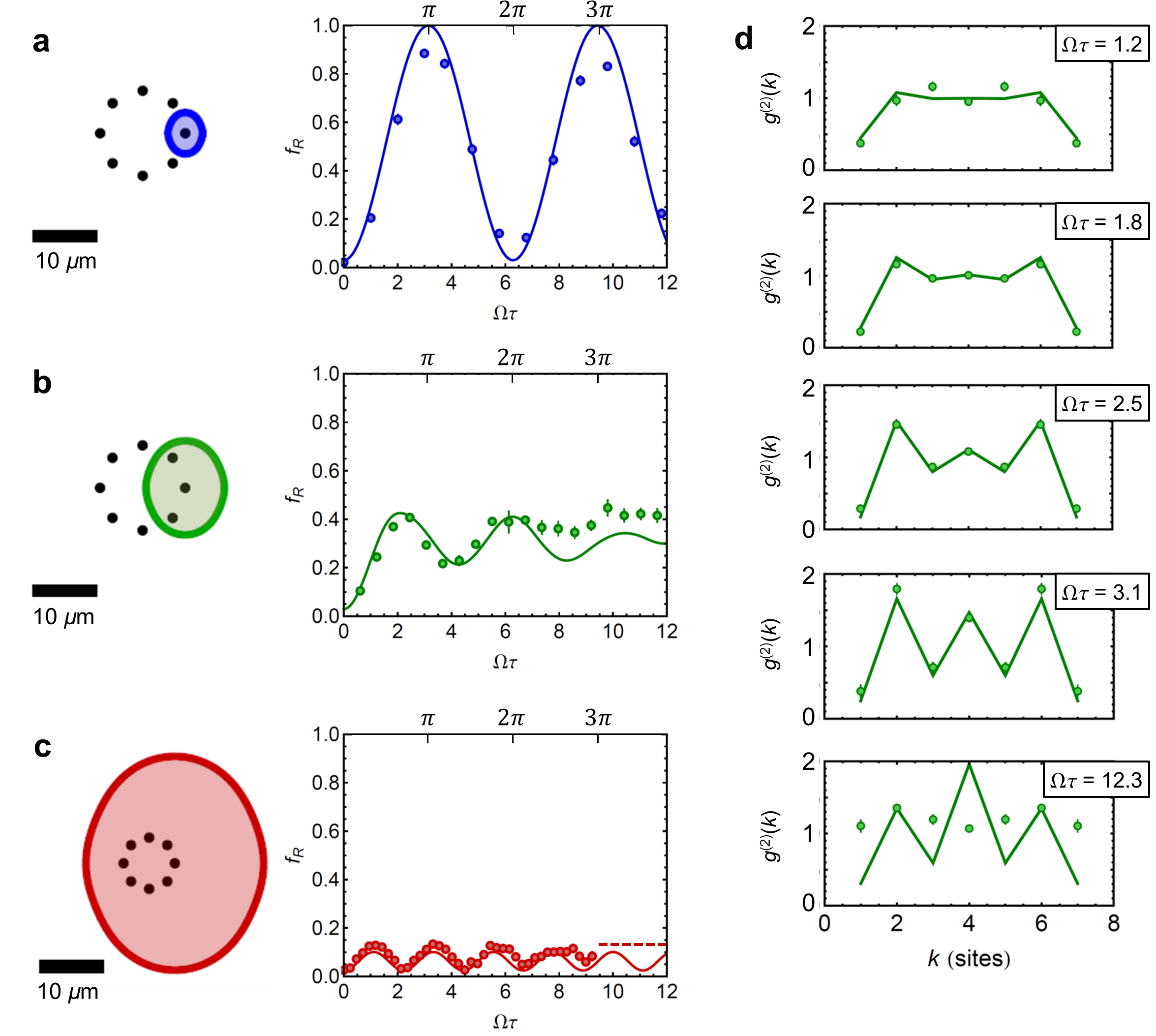}
\caption{\small {\bf Tuning interactions in an 8-spin chain with PBC.} {\bf a}: Independent atoms ($R_{\rm b}<a$). The Rydberg fraction $f_{\rm R}$ oscillates between $\simeq 0$ and $\simeq 1$, with the single-atom Rabi frequency $\Omega$. {\bf b}: Strongly correlated regime ($R_{\rm b}\simeq 1.5a$). The Rydberg fraction shows an oscillatory behavior involving several frequencies.  {\bf c}: Fully blockaded regime: $f_{\rm R}$ oscillates at $\sqrt{N}\Omega$, and reaches a maximum of $1/N$ (dashed line). {\bf d}: The Rydberg-Rydberg pair correlation function, for the parameters of {\bf b}, is shown for increasing values of $\Omega\tau$. In all plots, the solid lines are obtained by numerically solving the time-dependent Sch\"odinger equation, and then including detection errors ($\varepsilon =3\%$). Error bars (often smaller than symbol size) denote s.e.m. The shaded ellipsoids illustrate the (anisotropic) blockade volume.}
\label{fig:fig3}
\end{figure}

This versatility is illustrated in Fig.~\ref{fig:fig3}, where we use a fully loaded ring-shaped array of $N=8$ traps, thus realizing a small spin chain with periodic boundary conditions (PBC). By varying both $a$ and $n$, we tune the system all the way from independent atoms ($\alpha\ll 1$), where each atom undergoes a Rabi oscillation at frequency $\Omega$, resulting in a \emph{Rydberg fraction} $f_{\rm R}$ (defined as the average number of Rydberg excitations divided by $N$) periodically reaching $\simeq 1$ (Fig.~\ref{fig:fig3}a), to a fully blockaded array ($\alpha\gg 1$, Fig.~\ref{fig:fig3}c) characterized by collective oscillations  at frequency $\sqrt{N}\Omega$ and a maximum $f_{\rm R}=1/N$. In between (Fig.~\ref{fig:fig3}b, where $\alpha\simeq 1.5$), the evolution of $f_{\rm R}(\tau)$ shows oscillations resulting from the beating of the incommensurate eigen-frequencies of the many-body Hamiltonian~(1). Our system allows us to detect the state of each atom, and thus to measure correlation functions. Figure~\ref{fig:fig3}d shows the dynamics of the Rydberg-Rydberg pair correlation function
\begin{equation}
g^{(2)}(k)=\frac{1}{N_{\rm t}}\sum_{i}\frac{\langle n_{i} n_{i+k}\rangle}{\langle n_{i} \rangle \langle n_{i+k} \rangle}.
\label{eq:g2}
\end{equation}
The averaging over all traps does not wash out correlations despite the fact that the system is not fully invariant by translation (Methods). We observe a strong suppression of $g^{(2)}(k)$ for $k=1$ and $k=7$, i.e. a clear signature of nearest-neighbor blockade. For some times (see e.g. $\Omega\tau=3.1$), we observe an antiferromagnetic-like staggered correlation function, while the average density is uniform (Methods).

The solid lines in all panels of Figure~\ref{fig:fig3} are obtained by solving the Schr\"odinger equation governed by (\ref{eq:ising}) using the independently measured experimental parameters, and then including the effects of the finite detection errors $\varepsilon$ (Methods). One observes an overall agreement with the data, although some small discrepancies can clearly be noticed, especially at longer times. We attribute them to the Zeeman structure of Rydberg $D$-states, which is not taken into account in our modeling by a spin-$1/2$: for $\theta\neq 0$, the van der Waals interaction couples $\ket{r}$ to other Zeeman states, leading to a slow increase in the number of excitations (Methods).

\begin{figure}[t!]
\centering
\includegraphics[width=85mm]{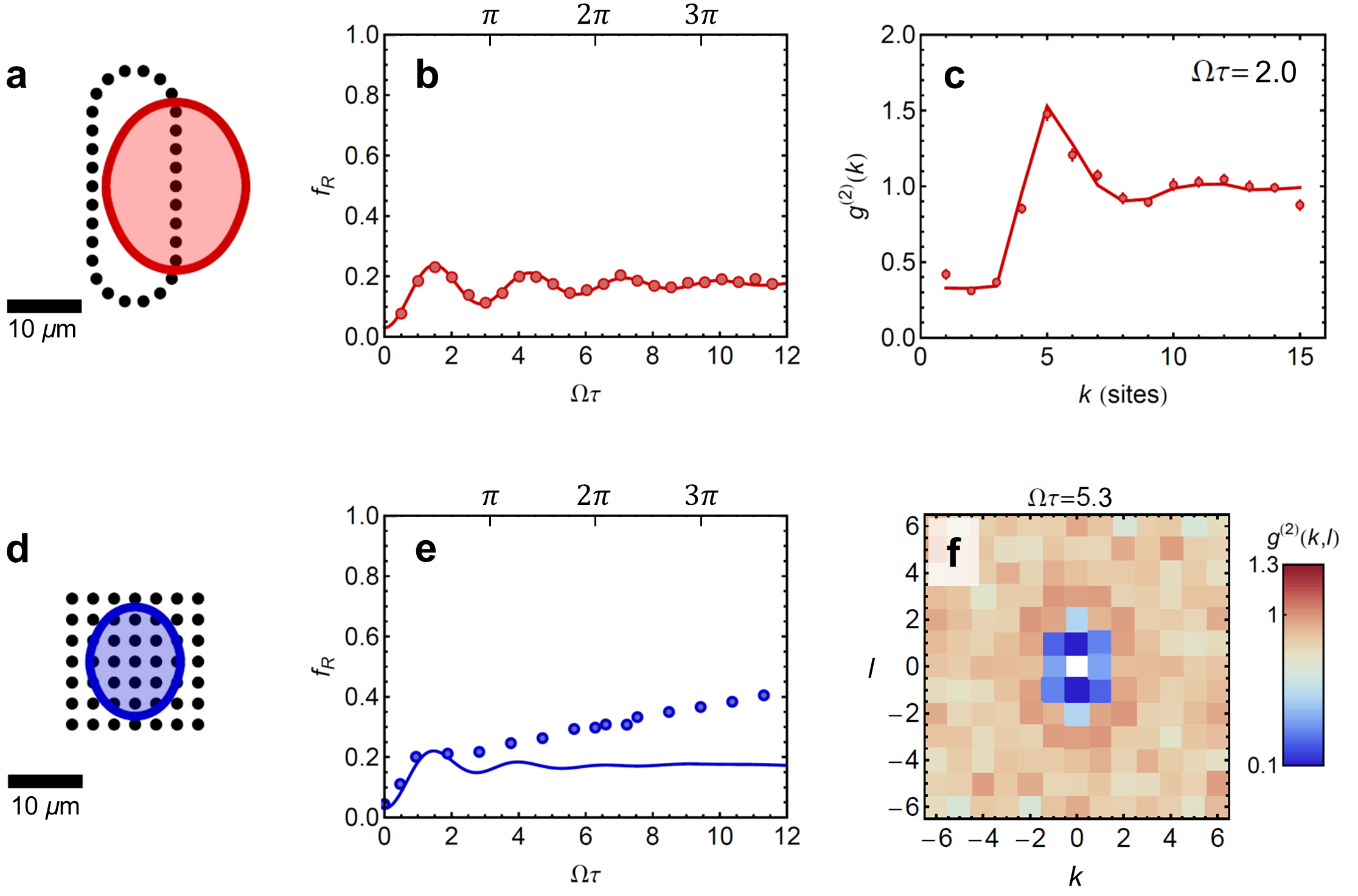}
\caption{\small {\bf Ising dynamics in large spin ensembles.} {\bf a}: Racetrack-shaped array with $N_{\rm t}=30$ traps, loaded with $N = 20\pm1.5$ atoms. The blockade radius $R_{\rm b}$ is about $4.3a$ (shaded ellipsoid). {\bf b}: Time evolution of the Rydberg fraction $f_{\rm R}$. {\bf c}: Rydberg pair correlation function $g^{(2)}(k)$  for $\Omega\tau\simeq 2.0$, showing a strong depletion for $k<R_{\rm b}$, and contrasted oscillations around the asymptotic value $1$. Error bars (most of the time smaller than symbol size) denote the s.e.m. Solid lines are the simulation results without any adjustable parameters. {\bf d}: Square array of $7\times 7$ traps loaded with $N= 28\pm1.6$ atoms. The blockade radius is about $2.6a$. {\bf e}: Evolution of $f_{\rm R}$. {\bf f}: Rydberg-Rydberg correlation function $g^{(2)}(k,l)$ for $\Omega\tau=5.3$.  }
\label{fig:fig4}
\end{figure}

We now study two systems containing a larger number of atoms. We first consider a one-dimensional spin chain with PBC comprising $N_{\rm t}=30$ traps and partially loaded with $N=20\pm 1.5$ atoms (Fig.~\ref{fig:fig4}a; we have checked that the $67\%$ filling fraction does not change qualitatively the physics as compared to a perfect filling, see Methods).  Its `racetrack' shape was chosen to optimize homogeneity of the Rabi frequency over the array. We chose parameters such that $\alpha\simeq4.3(1)$. The Rydberg fraction $f_{\rm R}(\tau)$ shows initial oscillations before reaching a steady state (Fig.~\ref{fig:fig4}b) due to the dephasing of the many incommensurate eigenfrequencies of (1) for this large value of $N$. The pair correlation function (shown in Fig.~\ref{fig:fig4}c for $\Omega\tau\simeq 2.0$) is strongly suppressed for $k<\alpha$, as expected from blockade physics, before oscillating towards the asymptotic value $g^{(2)}(k\gg\alpha)=1$\bibl{17,18}. A similar liquid-like correlation function has been observed in two dimensions\bibl{19}. The solid lines in Fig.~\ref{fig:fig4}b,c give the result of a full numerical simulation, without any adjustable parameters. Here the agreement with the spin-$1/2$ model is excellent, as many atom pairs are aligned along the quantization axis, thus making the effects of the anisotropy small. We included the finite value of $\varepsilon$, which has a strong effect on the pair correlations for $k<\alpha$ as $g^{(2)}(k)$ increases from 0 to $2\varepsilon/f_{\rm R}$ (Methods).

As a final setting, we use a $N_{\rm t}=7\times7$ two-dimensional square array (Fig.~\ref{fig:fig4}d), loaded with $N=28\pm 1.6$ atoms ($57\%$ filling), for $\alpha=2.6$. The dynamics of $f_{\rm R}$ now appears monotonous, without the initial oscillations seen above for smaller systems (Fig.~\ref{fig:fig4}e). This suggests that already with $N\sim 30 $ atoms, the behavior of the system is close to the many-body one observed in large ensembles\bibl{20} with a fast initial  rise of the Rydberg fraction, before it saturates. The simulation captures well the initial rise of $f_{\rm R}$, but does not reproduce the slow increase observed at long times, which we attribute again to multilevel effects (that are indeed expected to be strong in this array where the internuclear axes of many pairs lie at a large $\theta$). Figure~\ref{fig:fig4}f shows the two-dimensional Rydberg-Rydberg correlation function 
\begin{equation}
g^{(2)}(k,l)=\frac{1}{N_{\rm t}}\sum_{i,j}\frac{\langle n_{i,j} n_{i+k,j+l}\rangle}{\langle n_{i,j} \rangle \langle n_{i+k,j+l} \rangle}
\label{eq:corr}
\end{equation}
where $n_{i,j}$ refers to the site with coordinates $(ia,ja)$. Although the system has open boundaries and thus does not show translational invariance, the averaging over the traps in Eq.~(\ref{eq:corr}) does not wash out correlations as $R_{\rm b}$ is small compared to the system size. We observe a clear depletion of the correlation function close to the origin due to blockade. The anisotropy of the interaction is visible, as the depletion region is elliptical, with a flattening close to $1.2$. The full time evolution of the correlation function is shown in the online Methods.

The wide tunability of geometry and interactions demonstrated here opens fascinating perspectives for investigating the physics of spin systems with tens of particles. Our platform, especially when combined with quasi-deterministic loading of optical tweezers as demonstrated recently\bibl{21,22}, will be ideally suited for studying the transition from few- to many-body physics\bibl{23}, thermalization in strongly interacting closed quantum systems\bibl{24}, or the dynamical emergence of entanglement following a quantum quench\bibl{25}. Using resonant dipole-dipole interactions between different Rydberg states\bibl{26}, one can also implement XY Hamiltonians with long-range couplings\bibl{27}. Finally, exploiting the Zeeman structure of Rydberg states holds the promise of implementing more complex Hamiltonians, to explore for instance the physics of higher spins\bibl{28}, or realize topological insulators\bibl{29}.

\noindent{\bf Online Content:} Methods, along with any additional Extended Data display items, are available in the online version of the paper; references unique to these sections appear only in the online paper.

\section*{References and notes}

\begin{enumerate}

\item
Auerbach, A.
\emph{Interacting Electrons and Quantum Magnetism} (Springer-Verlag, New York, 1994).

\item
Georgescu, I. M., Ashhab, S. \& Nori, F.
\emph{Quantum simulation}. 
{Rev. Mod. Phys. {\bf 86}, 153 (2014)}.

\item
Bloch, I., Dalibard, J., \& Nascimb\`ene, S. 
\emph{Quantum simulations with ultracold quantum gases}.
{Nature Phys. {\bf 8}, 267 (2012)}.

\item
Yan, B., Moses, S.A., Gadway, B., Covey, J.P., Hazzard,  K.R.A., Rey, A.M., Jin, D.S., \& Ye, J. 
\emph{Observation of dipolar spin-exchange interactions with lattice-confined polar molecules}.
{Nature {\bf 501}, 521 (2013)}.

\item
Blatt, R. \& Roos, C.F.
\emph{Quantum simulations with trapped ions}.
{Nature Phys. {\bf 8}, 277 (2012)}.

\item
Glaetzle, A.W., Dalmonte, M., Nath, R., Rousochatzakis, I., Moessner, R., \& Zoller, P.
\emph{Quantum Spin-Ice and Dimer Models with Rydberg Atoms}.
{Phys. Rev. X {\bf 4}, 041037 (2014)}.

\item
Vermersch, B., Glaetzle, A.W., \& Zoller, P.
\emph{Magic distances in the blockade mechanism of Rydberg P and D states}.
{Phys. Rev. A {\bf 91}, 023411 (2015)}.

\item
Tresp, C., Bienias, P., Weber, S., Gorniaczyk, H., Mirgorodskiy, I., B\"uchler, H.P., \& Hofferberth, S.
\emph{Dipolar dephasing of Rydberg D-state polaritons}.
{Phys. Rev. Lett. {\bf 115}, 083602 (2015)}.
 
\item
Saffman, S., Walker, T.G., \& M{\o}lmer, K.
\emph{Quantum information with Rydberg atoms}.
{Rev. Mod. Phys. {\bf 82}, 2313 (2010)}.

\item
Weimer, H., M\"uller, M., Lesanovsky, I., Zoller, P., \& B\"uchler, H.P.
\emph{A Rydberg quantum simulator}.
{Nat. Phys. {\bf 6}, 382 (2010)}.

\item
Barredo, D., Ravets, S., Labuhn, H., B\'eguin, L., Vernier, A., Nogrette, F., Lahaye, T. \& Browaeys A.
\emph{Demonstration of a Strong Rydberg Blockade in Three-Atom Systems with Anisotropic Interactions}.
{Phys. Rev. Lett. {\bf 112}, 183002 (2014)}.

\item
B\'eguin, L., Vernier, A., Chicireanu, R., Lahaye, T., \&  Browaeys, A.
\emph{Direct Measurement of the van der Waals Interaction between Two Rydberg Atoms}.
{Phys. Rev. Lett. {\bf 110}, 263201 (2013)}.

\item
Nogrette, F., Labuhn, H., Ravets, S., Barredo, D., B\'eguin, L., Vernier, A., Lahaye, T. \&. Browaeys, A.
\emph{Single-Atom Trapping in Holographic 2D Arrays of Microtraps with Arbitrary Geometries}.
{Phys. Rev. X {\bf 4}, 021034 (2014)}.

\item
Dudin, Y.O., Li, L., Bariani, F., \& Kuzmich, A.
\emph{Observation of coherent many-body Rabi oscillations}.
{Nat. Phys. {\bf 8}, 790 (2012)}.

\item
Ebert, M., Gill, A., Gibbons, M., Zhang, X., Saffman, M., \& Walker, T.G.
\emph{Atomic Fock state preparation using Rydberg blockade}.
{Phys. Rev. Lett. {\bf 112}, 043602 (2014)}.

\item
Zeiher, J., Schau\ss, P., Hild, S., Macr\`i, T., Bloch, I., \& Gross, C.
\emph{Microscopic Characterization of Scalable Coherent Rydberg Superatoms}.
{Phys. Rev. X {\bf 5}, 031015 (2015)}.

\item
Ates, C., \& Lesanovsky, I. 
\emph{Entropic enhancement of spatial correlations in a laser-driven Rydberg gas}.
{Phys. Rev. A {\bf 86} 013408 (2012)}. 

\item
Petrosyan, D., H\"oning, M., \& Fleischhauer, M.,
\emph{Spatial correlations of Rydberg excitations in optically driven atomic ensembles}.
{Phys. Rev. A {\bf 87}, 053414 (2013)}.

\item
Schauss, P., Cheneau, M., Endres, M., Fukuhara, T., Hild, S., Omran, A., Pohl, T., Gross, C., Kuhr, S., \& Bloch, I.
\emph{Observation of Spatially Ordered Structures in a Two-Dimensional Rydberg Gas}.
{Nature {\bf 491}, 87 (2012)}.

\item
L\"ow, R., Weimer, H., Krohn, U., Heidemann, R., Bendkowsky, V., Butscher, B., B\"uchler, H.P., \& Pfau, T.
\emph{Universal scaling in a strongly interacting Rydberg gas}.
{Phys. Rev. A {\bf 80}, 033422 (2009)}.

\item
Lester, B.J., Luick, N., Kaufman, A.M., Reynolds, C.M., \& Regal, C.A.
\emph{Rapid production of uniformly-filled arrays of neutral atoms}.
{Phys. Rev. Lett. {\bf 115}, 073003 (2015)}.

\item
Fung, Y.H., \&  Andersen, M.F.
\emph{Efficient collisional blockade loading of single atom into a tight microtrap}.
{New J. Phys. {\bf 17}, 073011 (2015)}.

\item
Gaj, A., Krupp, A.T., Balewski, J.B., L\"ow, R., Hofferberth, S., \& Pfau, T.
\emph{From molecular spectra to a density shift in dense Rydberg gases}.
{Nature Comm. {\bf 5}, 4546 (2014)}.

\item
Ates, C., Garrahan, J.P., \& Lesanovsky, I.
\emph{Thermalization of a Strongly Interacting Closed Spin System: From Coherent Many-Body Dynamics to a Fokker-Planck Equation}.
{Phys. Rev. Lett. {\bf 108}, 110603 (2012)}.

\item
Hazzard, K.R.A., van den Worm, M., Foss-Feig, M.,  Manmana, S.R., Dalla Torre, E.G., Pfau,  T., Kastner, M. \& Rey, A.M.
\emph{Quantum correlations and entanglement in far-from-equilibrium spin systems}.
{Phys. Rev. A {\bf 90}, 063622 (2014)}.

\item
Barredo, D., Labuhn, H., Ravets, S., Lahaye, T., Browaeys, A., \& Adams, C.S. 
\emph{Coherent Excitation Transfer in a Spin Chain of Three Rydberg Atoms}.
{Phys. Rev. Lett. {\bf 114}, 113002 (2015)}.

\item
Hauke, P., Cucchietti,  F.M., M\"uller-Hermes, A., Ba\~{n}uls, M.-C., Cirac,  J.I., \& Lewenstein, M. 
\emph{Complete devil's staircase and crystal-superfluid transitions in a dipolar XXZ spin chain: a trapped ion quantum simulation}.
{New J. Phys. {\bf 12}, 113037 (2010)}.

\item
Senko, C., Richerme, P., Smith, J., Lee, A., Cohen, I., Retzker, A., \& Monroe, C. 
\emph{Realization of a Quantum Integer-Spin Chain with Controllable Interactions}.  
{Phys. Rev. X {\bf 5}, 021026 (2015)}.

\item
Peter, D., Yao, N.Y., Lang, N., Huber, S.D., Lukin, M.D., \& B\"uchler, H.P.
\emph{Topological bands with a Chern number $C=2$ by dipolar exchange interactions}.
{Phys. Rev. A {\bf 91}, 053617 (2015)}.

\end{enumerate}

\noindent
{\bf Acknowledgments:} We thank H. Busche for contributions in the early stages of the experiment, I. Lesanovsky, H.P. B\"uchler, and T. Pohl for useful discussions, and Y. Sortais for a careful reading of the manuscript. This work benefited from financial support by the EU [FET-Open Xtrack Project HAIRS, H2020 FET-PROACT Project RySQ, and EU Marie-Curie Program ITN COHERENCE FP7-PEOPLE-2010-ITN-265031 (H.L.)], by the `PALM' Labex (project QUANTICA) and by the Region \^Ile-de-France in the framework of DIM Nano-K.

\noindent
{\bf Author Contributions:}
H.L. and D.B. contributed equally to this work. H.L., D.B., S.R. and S.d.L. carried out the experiments, T.M. did the numerical simulations, T.L. and A.B. supervised the work. All authors contributed to the design of the experiments and to the data analysis. The manuscript was written by T.L. with input from all authors. 

\noindent
{\bf Author information:}
The authors have no competing financial interests. Correspondence and requests for material should be addressed to T.L. (thierry.lahaye@institutoptique.fr).

\newpage
\onecolumngrid
\section*{Methods}

\setcounter{figure}{0}

\renewcommand{\figurename}{\textbf{\small Extended Data Figure}}
\renewcommand{\tablename}{\textbf{\small Extended Data Table}}

\makeatletter
\makeatletter \renewcommand{\fnum@figure}
{\figurename~\textbf{\small\thefigure ${\bs |}$}}
\makeatother

\makeatletter
\makeatletter \renewcommand{\fnum@table}
{\tablename~\textbf{\small\thetable ${\bs |}$}}
\makeatother

\setcounter{equation}{3} 

\subsection*{Loading of trap arrays}

In the single-atom loading regime of optical microtraps, the probability to have a given trap filled with a single atom is $p\simeq 1/2$. Therefore, when we monitor the number of loaded traps in view of triggering the experiment, $N$ fluctuates in time around a mean value $N_{\rm t}/2$, with fluctuations $\sim\sqrt{N_{\rm t}}$. 

\begin{figure}[t!]
\centering
\includegraphics[width=0.7\linewidth]{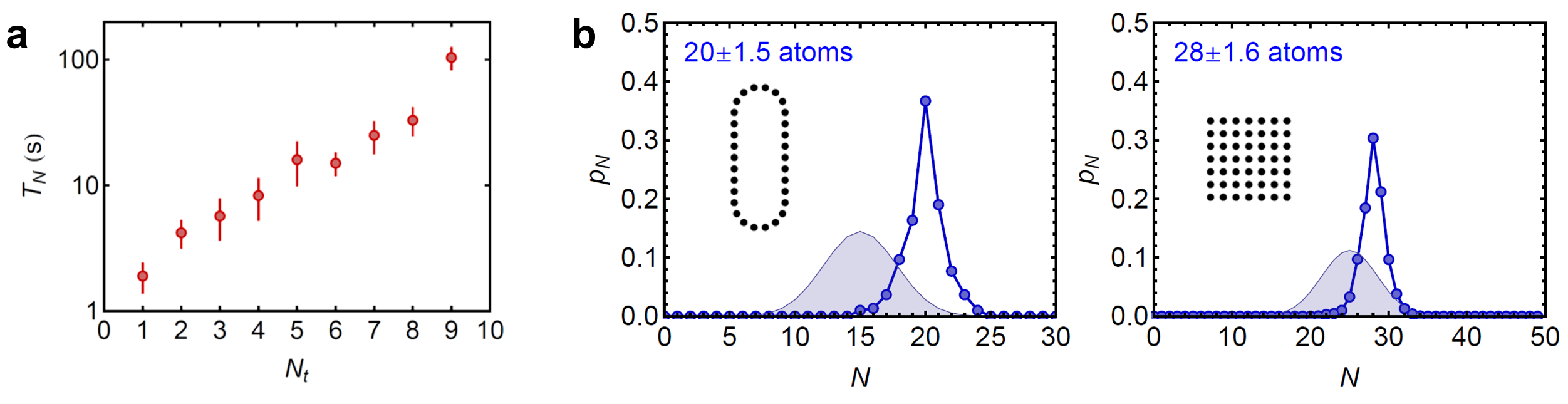}
\caption{\small {\bf Full and partial loading of arrays.} {\bf a}: Average triggering time $T_N$ when the triggering criterion is set to $N=N_{\rm t}$: achieving full loading requires an exponentially long time, limiting in practice the method to $N_{\rm t}\leqslant 9$. The triggering times can vary substantially depending on the density of the magneto-optical trap used to load the array, and the data points shown here correspond to typical conditions used for the data of main text. {\bf b}: Distribution of the number of loaded traps in the partially loaded regime for the 30-trap `racetrack' and the 49-trap square array (blue dots). The shaded distributions correspond to what would be observed with random triggering. }
\label{fig:loading}
\end{figure}

When the number of traps is small, we can impose, as the triggering criterion, to wait until all traps are filled. The average triggering time $T_N$ then increases exponentially with $N$, as can be seen in Extended Data Figure~\ref{fig:loading}a. We used this `full-loading mode' for the data of Fig.~1 ($1\leqslant N\leqslant9$) and Fig.~3 ($N=8$) of main text. This exponential scaling sets a practical limit of $N\sim9$ for fully loaded arrays. Already for $N=9$, the experimental duty cycle exceeds one minute.

Due to this, for larger $N_{\rm t}$ we use partially-loaded arrays. We set the triggering threshold in the tail of the binomial distribution of $N$, i.e. close to $N_{\rm t}/2+\sqrt{N_{\rm t}}$. This allows us to keep a fast repetition rate for the experiment, on the order of $1\;{\rm s}^{-1}$, enabling fast data collection. Extended Data Figure~\ref{fig:loading}b shows the distribution of loaded traps for the `racetrack' array  with $N_{\rm t}=30$ (respectively, for the $N_{\rm t}=7\times7$ square array), where we set the triggering condition to $N=20$ (resp. $N=30$). Using this triggering procedure, we thus end up with a narrow distribution of atom numbers $N=20\pm1.5$ (resp. $N=28\pm1.6$), corresponding to a filling fraction of 67\% (resp. 57\%), significantly above the average $N_{\rm t}/2$. These strongly subpoissonian distributions of atom numbers are such that the variation in $N$ from experiment to experiment has a negligible effect on the physics studied in Fig.~4 of main text; however, as for each experiment the initial configuration image is saved, one can if needed post-select experiments where an exact number of atoms was involved (this is how the data in Fig.~2 of main text for $N\geqslant 10$ were obtained). 
 
Recently, several experiments\bibl{21,22} demonstrated quasi-deterministic loading of single atoms in optical tweezers, reaching $p\sim 90\%$ using modified light-assisted collisions that lead to the loss of only one of the colliding atoms instead of both. A preliminary implementation of these ideas on our setup gave $p\sim 80\%$ for a single trap. In future work, by using such loading in combination with the real-time triggering based on the measured number of loaded traps, it seems realistic to reach, even in large arrays, filling fractions in excess of 0.9, i.e. approaching those obtained in quantum gas microscope experiments using Mott insulators.

\subsection*{Experimental parameters}

Extended Data Table~\ref{tab:tab1} summarizes the various values of the parameters of the arrays of traps and of the Rydberg states used for the data presented in the main text, and the resulting values of the dimensionless parameter $\alpha$. It illustrates the wide tunability offered by the system.

\begin{table}[b!]
\begin{center}
\begin{tabular}{lcccccccc}
\hline
             & \multicolumn{3}{c}{\bf Trap array parameters} & \multicolumn{4}{c}{\bf Rydberg state parameters}                         &          \\
           
{\bf Figure} & Spacing $a$     & $N_{\rm t}$   & $N$        & $n$         & Calculated  $C_6/h$     &  $\Omega/(2\pi)$    & $R_{\rm b}$     & $\alpha$ \\
             & $({\rm \mu m})$ &               &            &             & $({\rm GHz\,\mu m^6})$  &  ${\rm (MHz)}$      & $({\rm \mu m})$ &          \\
             \hline \hline
             &                 &               &            &             &                         &                     &                 &          \\
2a (full)    &  3.0            &  1--9         &$N_{\rm t}$ & 82          &    $-8.9\times 10^3$    &      $1.5$          &    $14$         &  $4.5$   \\
2a (partial) &  3.2            &  19           &  10--15    & 100         &    $-8.0\times 10^4$    &      $1.1$          &    $20$         &  $6.4$   \\
             &                 &               &            &             &                         &                     &                 &          \\
3a           &  6.3            & 8             & 8          & 54          &    $-6.7$               &      $1.6$          &    $4.0$        &  $0.63$  \\
3b           &  6.3            & 8             & 8          & 61          &   $-7.6\times 10^2$     &      $1.3$          &    $9.1$        &  $1.4$   \\
3c           &  3.8            & 8             & 8          & 100         &   $-8.0\times 10^4$     &      $0.95$         &   $21$          &  $5.5 $  \\
             &                 &               &            &             &                         &                     &                 &          \\
4a,b,c       &  3.1            & 30            & $20\pm 1.5$& 79          &  $-6.0\times 10^3$      &      $1.0$          &    $13.5$       &  $4.3$   \\
4d,e,f       &  3.5            & 49            & $28\pm1.6 $& 61          &  $-7.6\times 10^2$      &      $1.4$          &    $9.1$        &  $2.6$   \\                          
\hline
  
\end{tabular} 
\caption{\small {\bf Experimental parameters used for the data presented in the main text.} Wide tuning of $\alpha=R_{\rm b}/a$, over one order of magnitude, is achieved by a combination of changes in $a$ and $n$ (while $\Omega$ is kept almost constant). }
\label{tab:tab1}
\end{center}
\end{table}

\subsection*{Finite detection errors} \label{sec:epsilon}

Our way to detect that a given atom has been excited to a Rydberg state relies on the fact that we do not detect fluorescence from the corresponding trap in the final configuration image. There is however a small probability $\varepsilon$ to lose an atom during the sequence, even if it was in the ground state, thus incorrectly inferring its excitation to a Rydberg state\bibl{11}. These `false positive' detection events affect the measured populations of the $N$-atom system. One can show that, if $P_{q}$ is the \emph{observed} probability to have $q$ Rydberg excitations, and $\tilde{P}_{p}$ the \emph{actual} probability to have $p$ Rydberg excitations, 
\begin{equation}
P_{q}=\sum_{p=0}^q \binom{N-p}{q-p}\varepsilon^{q-p}(1-\varepsilon)^{N-q}\tilde{P}_{p}.
\label{eq:eps}
\end{equation}
In principle, one can invert the above linear system relating the observed and actual probabilities\bibl{30}, to correct the experimental data for the detection errors. Here we have chosen on the contrary to show the uncorrected populations, and to include detection errors on the theoretical curves instead. 

\begin{figure}[t!]
\centering
\includegraphics[width=0.45\linewidth]{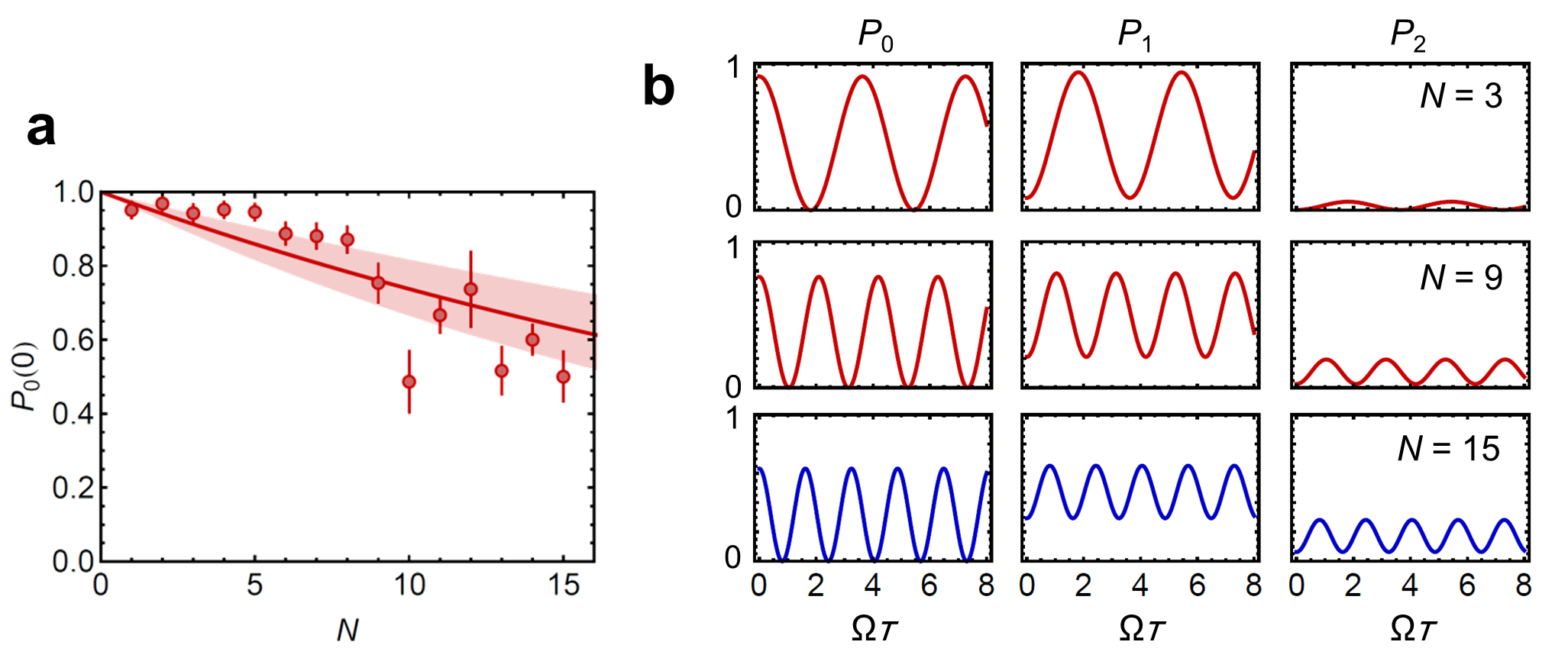}
\caption{\small {\bf Effect of detection errors.} {\bf a}: Experimental determination of $\varepsilon$. From the data of the full blockade experiments (Fig. 2 of main text), we plot the probability $P_0$ to recapture all $N$ atoms for $\tau=0$. The solid line is a fit to the expected dependence $(1-\varepsilon)^N$, giving $\varepsilon=3\%$ (the shaded area corresponds to $2\%<\varepsilon<4\%$). {\bf b}: Calculated probabilities to observe 0,1 or 2 excitations assuming a perfect blockade and $\varepsilon=3\%$, for atom numbers $N=3,9,15$.  }
\label{fig:epsilon}
\end{figure}

In order to determine the experimental value of $\varepsilon$, we use the initial datapoints ($\tau=0$) of the data of Fig.2 of main text. Since no Rydberg pulse is sent, we have $\tilde P_0=1$, and from (\ref{eq:eps}) the observed probability $P_0(\tau=0)$ reads $(1-\varepsilon)^N$. Extended Data Figure~\ref{fig:epsilon}a shows the variation of $P_0(0)$ as a function of $N$, together with a fit which allows us to extract $\varepsilon=(3\pm1)\%$, the value we use for the theoretical curves in the main text (see below).

Extended Data Figure~\ref{fig:epsilon}b shows the effect of this finite value of $\varepsilon$ on the probabilities $P_0$, $P_1$ and $P_2$ in the full blockade regime, for atom numbers $N=3,9,15$, clearly illustrating that the `false positive' detection events (i) yield non-zero (and increasing with $N$) double excitation probabilities (that oscillate in phase with $P_1$) (ii) multiply the amplitude of $P_0$ by a factor $(1-\varepsilon)^N$ and (iii) reduce the contrast of the $P_1$ oscillations. Globally, the experimental data (see Extended Data Figure~\ref{fig:block_som}) shows these features, superimposed with other imperfections such as damping, not related to the finite value of $\varepsilon$.

Finally, let us mention the effect of the detection errors on the correlation functions. In the fully blockaded region $k<\alpha$, one ideally expects a vanishing $g^{(2)}$ for $\varepsilon=0$. However, to lowest order in $\varepsilon$, this value is increased substantially (see e.g. Fig. 4c of main text) to $2\varepsilon / f_{\rm R}$ where $f_{\rm R}$ is the Rydberg fraction. Indeed, $g^{(2)}(k=1)$ is given by an average of quantities of the form $\langle n_i n_{i+1}\rangle/(\langle n_i\rangle \langle n_{i+1}\rangle)$. For $\varepsilon=0$, the numerator vanishes due to blockade; the only possibility to have a non-zero value comes from detection errors. To lowest order in $\varepsilon$, the probability to get a nonzero value for $n_i n_{i+1}$ is that either atom $i$ is in $\ket{r}$ (probability $f_{\rm R}$) and atom $i+1$ is lost (probability $\varepsilon$), or vice-versa. This results in a value $2\varepsilon f_{\rm R}$ for the numerator, while for the denominator we can use the zeroth-order values $\langle n_i\rangle=\langle n_{i+1}\rangle=f_{\rm R}$, thus giving $g^{2}(1)\simeq 2\varepsilon / f_{\rm R}$, which experimentally can be as large as 0.5.

\subsection*{Supplementary experimental data}

\begin{figure}[t!]
\centering
\includegraphics[width=0.8\linewidth]{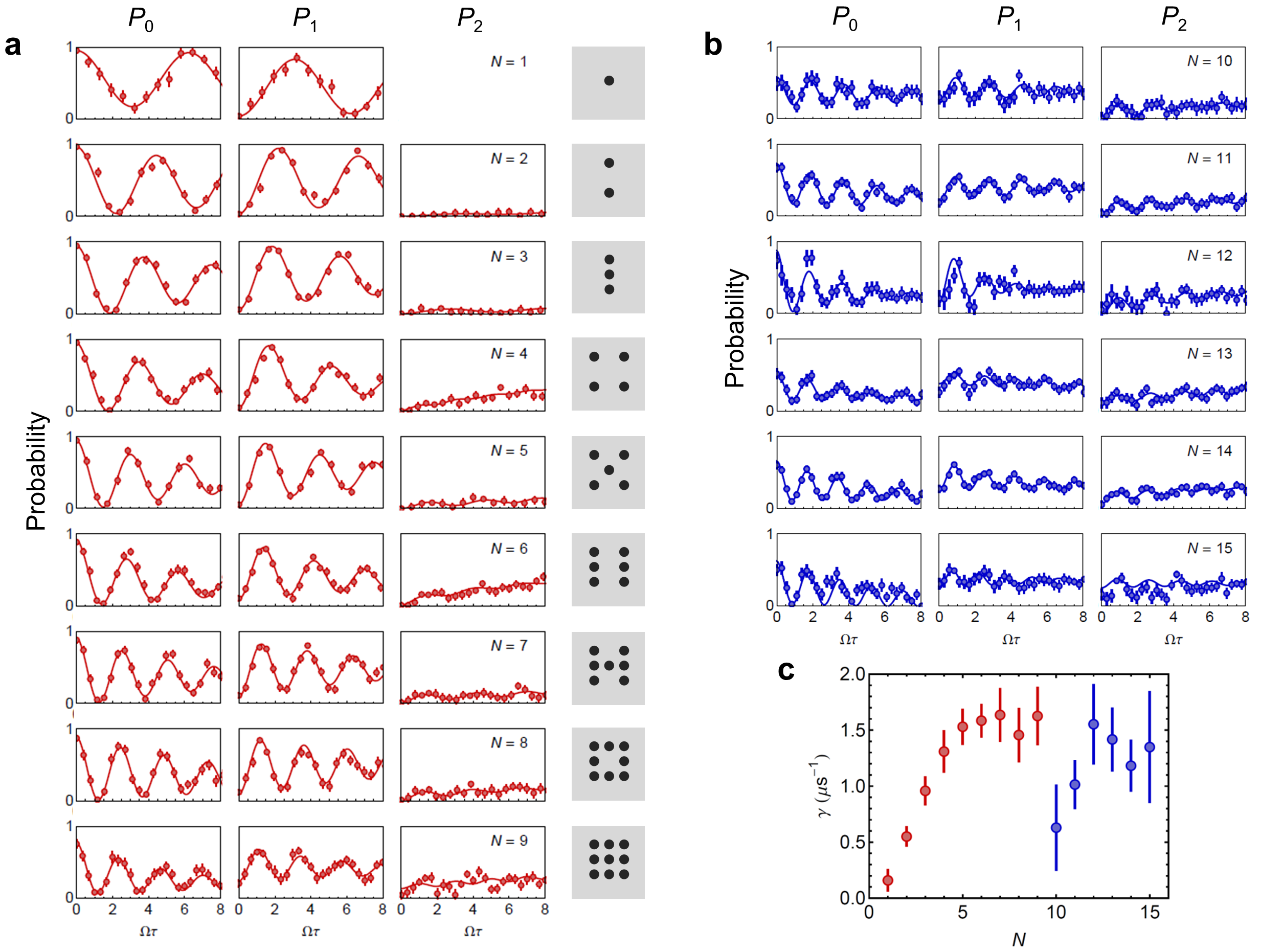}
\caption{\small {\bf Full dataset for the Rydberg blockade data.} {\bf a}: Fully loaded arrays of 1 to 9 traps ($n=82$). {\bf b}: partially loaded array of $N_{\rm t}=19$ traps, containing from $N=10$ to $N=15$ atoms ($n=100$). The column on the left shows the probability $P_{\rm 0 }$ to recapture all atoms, the center column the probability $P_{\rm 1 }$ to lose just one atom out of $N$, and the column on the right the probability $P_{\rm 2 }$ to lose two atoms out of $N$. The solid lines are fits by (\ref{eq:asym}). {\bf c}: Damping rate $\gamma$ extracted from the $P_{\rm 0 }$ data as a function of the number of atoms in the array.}
\label{fig:block_som}
\end{figure}

\begin{figure}[t!]
\centering
\includegraphics[width=0.5\textwidth]{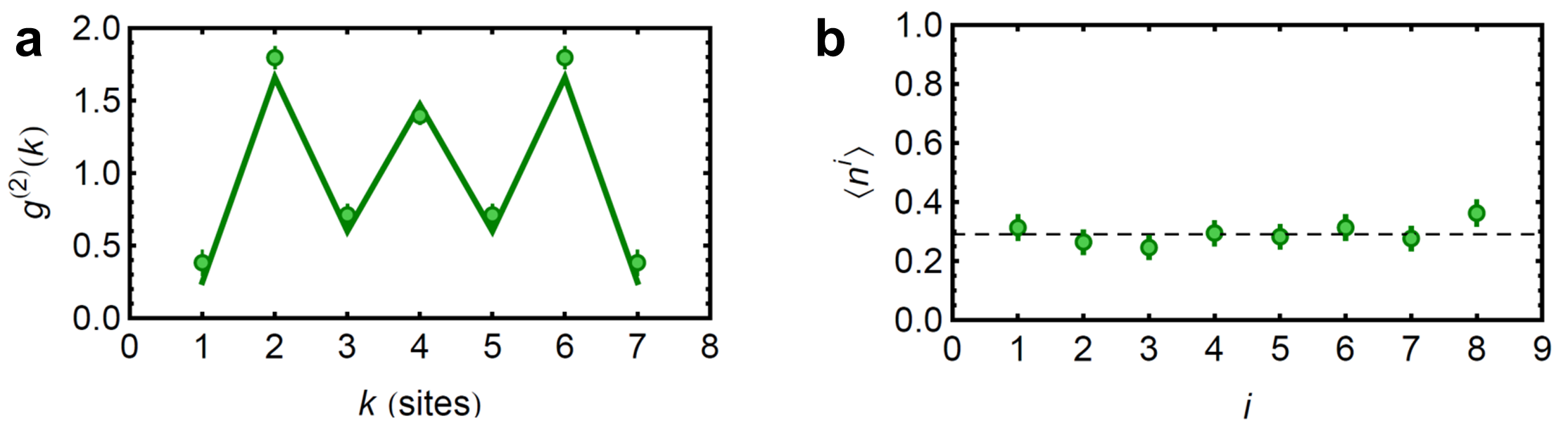}
\caption{\small {\bf Homogeneous excitation in the 8-atom ring.} {\bf a}: For $\Omega\tau=3.1$, we observe strongly contrasted oscillations in the pair correlation function $g^{2}(k)$. {\bf b}: The average \emph{density} of Rydberg excitations, however, is approximately the same on every site. The horizontal dashed line indicates the mean over all sites. }
\label{fig:som_ring_density}
\end{figure}

\begin{figure}[t!]
\centering
\includegraphics[width=0.7\linewidth]{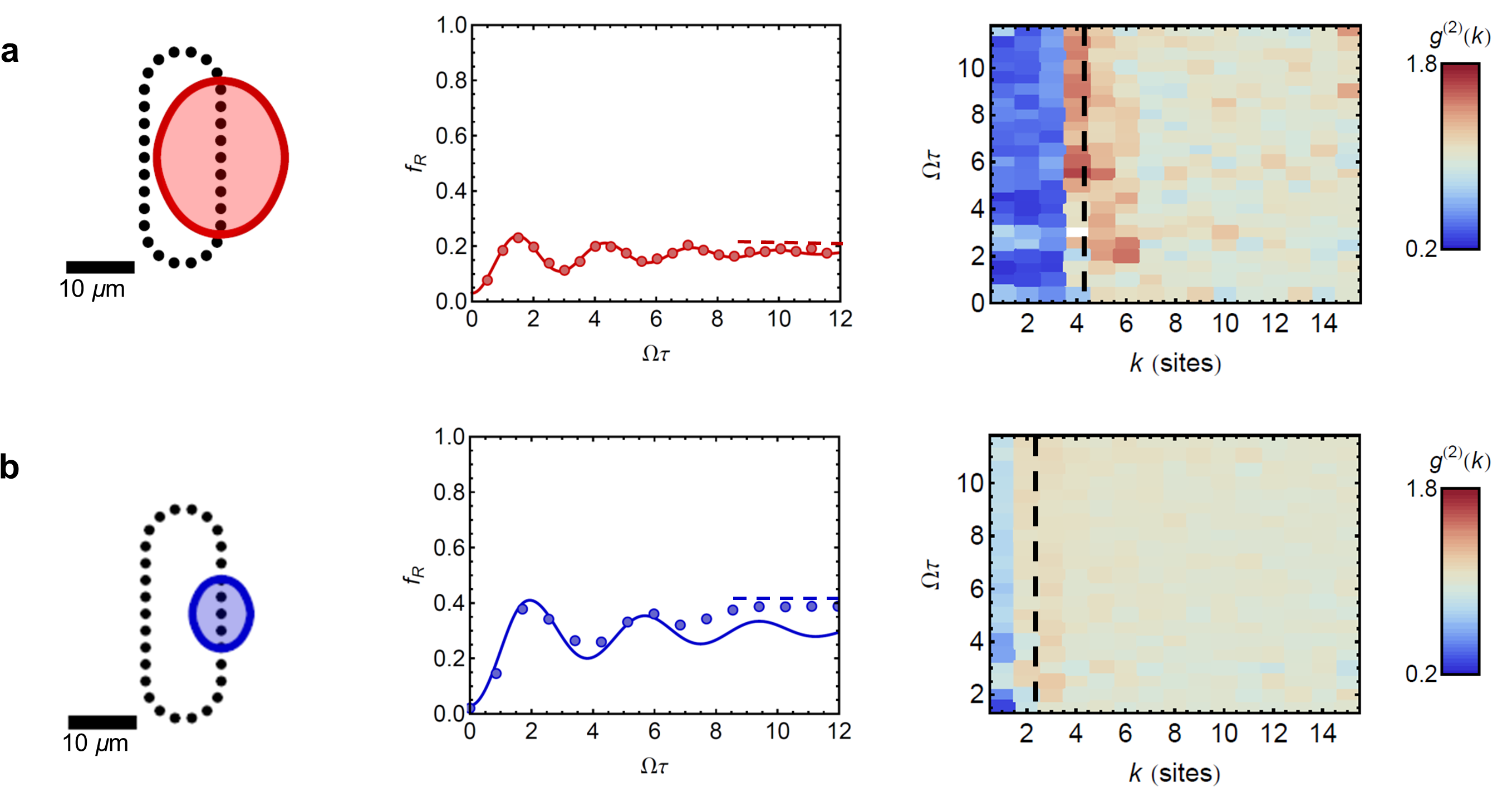}
\caption{\small {\bf Full time evolution of the correlation functions for the 30-trap, racetrack-shaped chain.} {\bf a}: Same as for Figure~3a,b,c of main text. The right panel is the time evolution of the pair correlation function, clearly showing that, for times longer than a few $\Omega^{-1}$, the pair correlation function does not evolve significantly anymore. The vertical dashed line indicates the value of the blockade radius.  {\bf b}: The principal quantum number is now $n=57$, and the Rabi frequency $\Omega=2\pi\times 1.7$~MHz, such that $R_{\rm b}=2.4a$. The central panel shows the time evolution of the Rydberg fraction, and the right panel the time evolution of the pair correlation function. For both {\bf a} and {\bf b}, $f_{\rm R}$ approaches, at long times, the close-packing limit $a/R_{\rm b}$ of hard rods of length $R_{\rm b}$ (dashed horizontal lines).}
\label{fig:figs1}
\end{figure}

\begin{figure}[t!]
\centering
\includegraphics[width=17cm]{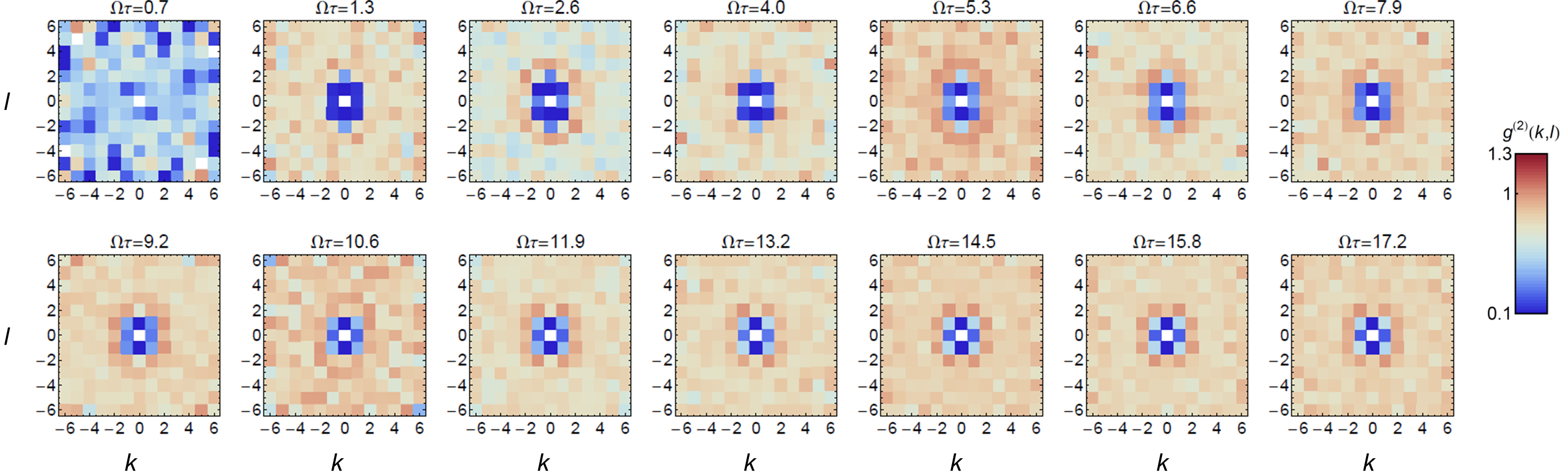}
\caption{\small {\bf  Full time evolution of the correlation function for the $7\times7$ square array}. One observes the blockaded region around $(k,l)=(0,0)$, with a slight flattening reflecting the anisotropy of the interaction. After a few $\Omega^{-1}$, the correlation function does not evolve any more. }
\label{fig:figs2}
\end{figure}

\emph{Full Rydberg blockade.---} Extended Data Figure~\ref{fig:block_som} shows additional data in the full blockade regime (Fig.~2 of main text). In Extended Data Figure~\ref{fig:block_som}a, the arrays of 1 to 9 traps are fully loaded, while in Extended Data Figure~\ref{fig:block_som}b, the 19-trap triangular array is partially loaded with 10 to 15 atoms. In both panels, the left column shows the time evolution of the probability $P_0$ to recapture all atoms at the end of the sequence, the middle column shows $P_1$, and the right column shows $P_2$. The points in Fig.~2a of main text corresponding to $N=8$ and $N=9$ in partially loaded arrays were taken in a similar configuration as for $N=10$ to $15$, but the array contained only $N_{\rm t}=17$ traps. The curves (not shown here) do not show any noticeable difference with other sets of data.
\begin{itemize}
\item We recognize the effects of the finite detection errors $\varepsilon\neq0$ on the amplitude and contrast of the collective oscillations discussed in section~\ref{sec:epsilon} above;
\item In addition, the oscillations exhibit some damping, which seems to increase with $N$. To quantify this, we fit the data by the function 
\begin{equation}
P(\tau)=a {\rm e}^{-\gamma \tau}\left(\cos^2(\Omega_N \tau/2)+b\right)+c,
\label{eq:asym}
\end{equation}
where $a,b,c,\gamma$ and $\Omega_N$ are adjustable parameters (solid lines). This functional form was chosen to account in a simple way for the asymmetry in the damping. Extended Data Figure~\ref{fig:block_som}c shows the damping rates $\gamma$, extracted from the probabilities $P_{\rm 0}$ as a function of $N$. We observe an initial increase in the damping rates, which then saturates above $N=5$. An increase with $N$ of the damping rate was observed in other similar blockade experiments\bibl{14-16}.
\item In addition, we observe that $P_2$ slowly increases over time for some specific values of $N$ (see in particular $N=4,6,9,13$), corresponding to particular geometries.  
\end{itemize}
We do not have a full understanding of these last two observations, but they may originate from the breaking of the blockade due to the Zeeman structure of the Rydberg states $nD_{3/2}$ (see discussion below).

\noindent\emph{8-atom ring.---} Extended Data Figure~\ref{fig:som_ring_density} shows that, within statistical fluctuations, the density of excitations on the 8-atom ring is homogeneous (this remains true at all times), and that the antiferromagnetic-like or crystal-like features obtained for some times, e.g. for  $\Omega\tau=3.1$, can only be observed in the correlation functions. This illustrates the interest of our setup in which spin chains with PBC can be realized  easily. On the contrary, in a 1D chain with open boundary conditions, `pinning' of the excitations at specific sites would occur due to edge effects.

\noindent\emph{Racetrack-shaped array.---} Extended Data Figure~\ref{fig:figs1}a shows the full evolution of the time correlation function for the data of Fig.~4abc of the main text ($R_{\rm b}=4.3a$). Extended Data Figure~\ref{fig:figs1}b  corresponds to the same settings except for the fact that one now has $R_{\rm b}=2.4a$.

\noindent\emph{Square array of $7\times7$ traps.---} 
Figure~\ref{fig:figs2} shows the full time evolution of the two-dimensional Rydberg-Rydberg correlation function $g^{(2)}(k,l)$ for the $7\times 7$ square lattice of Fig.~4def. Note that the two-dimensional pair correlation function is calculated using~(3), which implies that, due to the finite size of the array, the number of terms included in the sum decreases when $k,l$ increase. The normalization takes this variation into account.

\subsection*{Anisotropy of the interaction}

For a pair of atoms in a $nD_{3/2}$ Rydberg state with the internuclear axis not aligned with the quantization axis, the rigorous description of the van der Waals interaction requires to include all various Zeeman sublevels; the interaction then takes the form of a $16\times 16$ matrix. To keep the description of a system of $N$ atoms tractable, one can, in the blockade regime, define an effective, anisotropic van der Waals potential\bibl{7} reducing the previous matrix to a single scalar. For $nD_{3/2}$ states, the anisotropy reported in refs.~7,11 is well reproduced by the simple expression
\begin{equation}
V_{\rm eff}(r,\theta)=\frac{C_6(0)}{r^6}\left(\frac{1}{3}+\frac{2}{3}\cos^4{\theta}\right)
\label{eq:veff}
\end{equation}
with $\theta$ the angle between the quantization axis and the internuclear axis, giving a reduction by a factor of three in interaction strength when $\theta$ goes from $0$ to $\pi/2$. 

Due to the anisotropy in (\ref{eq:veff}), the shape of the blockade volume centered on a Rydberg atom is also anisotropic. However, due to the $r^6$-scaling of the interaction, the surface $r(\theta)$ defined by $V_{\rm eff}(r,\theta)=\hbar \Omega$ is quite well approximated by a prolate spheroid with an aspect ratio of $3^{1/6}\simeq 1.2$. In the figures of the main text, the shaded regions depicting the blockade volume have the polar equation $r(\theta)=R_{\rm b}\left(\frac{1}{3}+\frac{2}{3}\cos^4{\theta}\right)^{1/6}$.

\subsection*{Numerical simulation of the dynamics}

Our theoretical description of the system is based on the mapping of its dynamics into a pseudo-spin $1/2$ model with anisotropic long range interactions. We therefore neglect the rich Zeeman structure of the $nD_{3/2}$ states.  The numerical calculations rest on the solution of the Schr\"odinger equation for the Hamiltonian of Eq.~(1) 
of the main text in a reduced Hilbert space $\mathcal H$.  We first write the wave function $\left| \psi\right>$ of the system with $N$ atoms in  terms of states with fixed number of Rydberg excitations and ground state atoms,  which correspond to the eigenstates of the Hamiltonian with vanishing Rabi frequency $\Omega$\bibl{17,31}. Then the truncation procedure is based on two complementary steps: first we define the maximum number of Rydberg excitations $N_r^\text{max}$ that we include in our basis, second we eliminate those states which display excitations closer than a fixed distance $R_0$. Both $N_r^\text{max}$ and $R_0$ are adjusted to ensure the convergence of the dynamics.  For small samples (Fig.~3 of the main text) we performed simulations including all 256 basis states, whereas for the racetrack configurations we typically set $R_0$ smaller than the lattice constant but include up to $N_r^\text{math}=10$ excitations at most, reducing the dimension of $\mathcal H$ from $2^{20}\simeq 10^6$ to $\sum_{q=0}^{N_r^\text{max}}\binom{20}{q} \simeq 6\times10^5$. For the $7\times 7$ square array with 30 atoms, we set $R_0=1.3a$ (much smaller than the blockade radius $R_{\rm b}=2.6a$), thus reducing the dimension of $\mathcal H$ to $\simeq 3\times 10^6$ (the full Hilbert space is of  dimension $2^{30}\simeq 10^9$, and using only the truncation criterion on the number of excitations would reduce it to about $5\times10^7$, still intractably large). The Schr\"odinger equation within the truncated Hilbert space is then solved with standard split-step method for the two  non-commuting parts of the Hamiltonian of Eq.~(1) of the main text. All these calculations were repeated for several realizations of the loading of the arrays ($50$ realizations for the squared $7\times 7$ configurations and $200$ realizations for the case with fewer traps),  taking into account the anisotropic interparticle interaction of Eq.~(\ref{eq:veff}). The comparison with experimental data of the average fraction of excitations $f_{\rm R}=\sum_{q=0}^N qP_q/N$ is done by including the ``false positive'' detection events as described by Eq.~(\ref{eq:eps}).

The calculation of the $g^{(2)}(k)$ correlation function in Fig.3d and Fig.4c of the main text follows the definition of Eq.~(\ref{eq:s:g2}). However, contrarily to the calculation of the average fraction of the excitations it is not possible to derive an analytical formula for $g^{(2)}(k)$ to properly take into account the detection efficiency of Rydberg excitations (unless $k<\alpha$ as described in section S.1.3.).  Therefore we implement a standard Monte Carlo algorithm to perform the average of the correlation function over randomly generated configurations which are weighted in $g^{(2)}(k)$ with the initial (quantum) probability extracted from the real time dynamics of the Schr\"odinger equation.  For example the state $\left| r_{i} \, r_{j}\right>$ which contains $N_{\rm r}=2$ Rydberg excitations and amplitude $c_{{i\,j}}(t)$ can wrongly be dectected as the state $\left| r_{i} \, r_{j}  \, r_{q} \right>$ with probability $p = \varepsilon\, (1-\varepsilon)^{N-2}$.  If the latter state is generated from our sampling algorithm then its weight in the correlation function corresponds to $\left| c_{{i\,j}}(t)\right|^2$. Finally we average over several hundreds randomly generated configurations to obtain well converged results for the correlation function.

\subsection*{Effect of partial loading of large arrays on the observed dynamics}

Using the simulations described above, we explore to which extent the partial loading of our larger arrays may change the observed dynamics as compared to the ideal case of full loading.

\begin{figure}[t!]
\centering
\includegraphics[width=10cm]{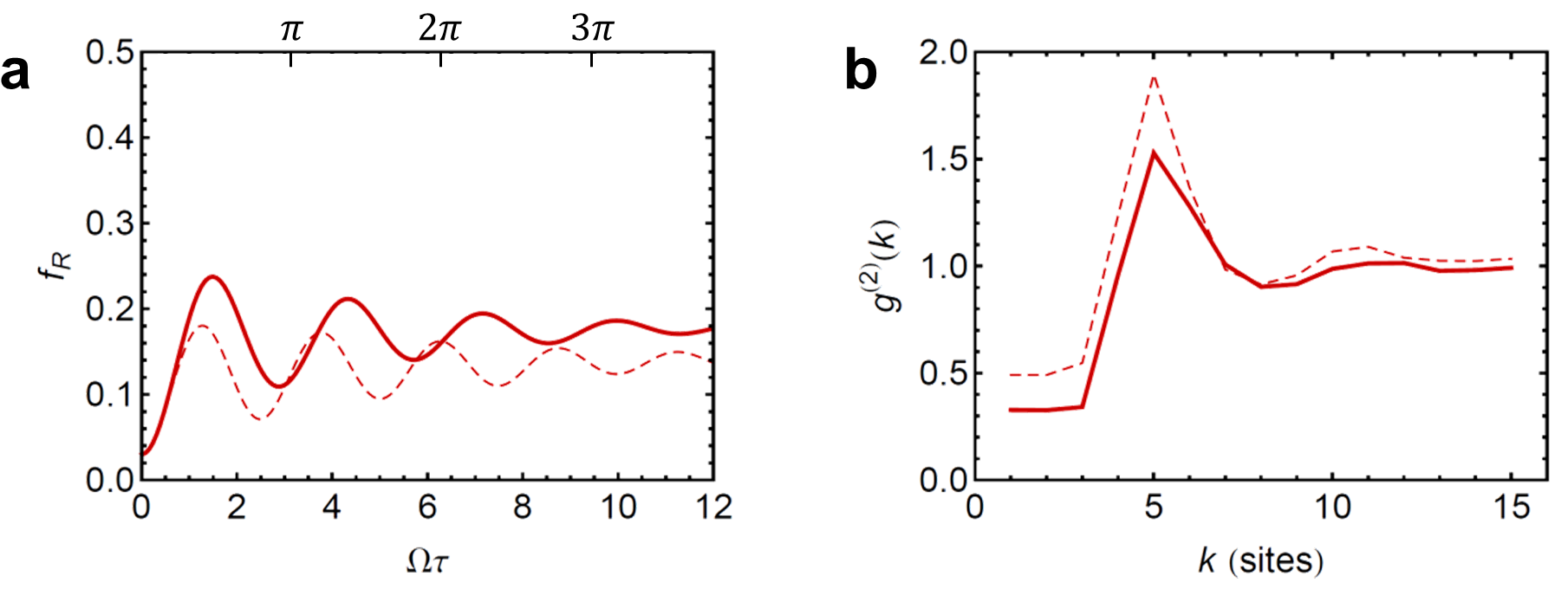}
\caption{\small {\bf Full versus partial loading for the dynamics and correlations in the case of Fig 4a,b,c of main text.} {\bf a}: Rydberg fraction as a function of time for the partially loaded (solid line) or fully loaded (thin dashed line) 30-trap array. {\bf b}: Pair correlation function $g^{(2)}(k)$ for $\Omega\tau\simeq2.0$, for the partially loaded (solid line) or fully loaded (thin dashed line) 30-trap array. In both cases, the effect of detection errors ($\varepsilon=3\%$) is included.}
\label{fig:som_partial}
\end{figure}

Extended Data Figure~\ref{fig:som_partial} shows, for the `racetrack' array of Fig. 4abc of main text, the results of simulations for the experimentally relevant case of partial loading (solid lines, filling fraction $\eta\simeq0.67$) and for the ideal, full loading case (thin dashed lines):
\begin{itemize}
\item Extended Data Figure~\ref{fig:som_partial}a shows the time evolution of the Rydberg fraction $f_{\rm R}$. The dynamics is qualitatively similar in the two situations, with initial oscillations that rapidly get damped due to the dephasing of the many incommensurate eigen-energies of the Hamiltonian. Quantitatively, the initial oscillations are faster in the fully loaded case: this is expected, as each blockade volume contains $1/\eta$ as many atoms, and thus, due to the scaling of the collective Rabi frequency with the number of atoms in a blockade volume, we expect an enhancement  of the oscillation frequency by $\sim \eta^{-1/2}\simeq 1.2$, close to what we observe. In the same way, the asymptotic Rydberg fraction when $\tau\to\infty$ is reduced by a factor close to the expected factor $\eta$.
\item Extended Data Figure~\ref{fig:som_partial}b shows the pair correlation function $g^{(2)}(k)$ for $\Omega\tau\simeq2.0$. Here again, the changes are moderate, although the oscillations of the correlation function for $k>\alpha$ would be slightly more contrasted for the fully loaded array.
\end{itemize}

Simulations for the other large array settings give similar results, allowing us to safely conclude that the partial loading of our largest arrays does not affect significantly the observed dynamics. This conclusion would be different for other types of experiments, for instance the transport of a spin excitation in the case of resonant-dipole-dipole interactions.

\subsection*{Approximative translational invariance}

\begin{figure}[b!]
\centering
\includegraphics[width=15cm]{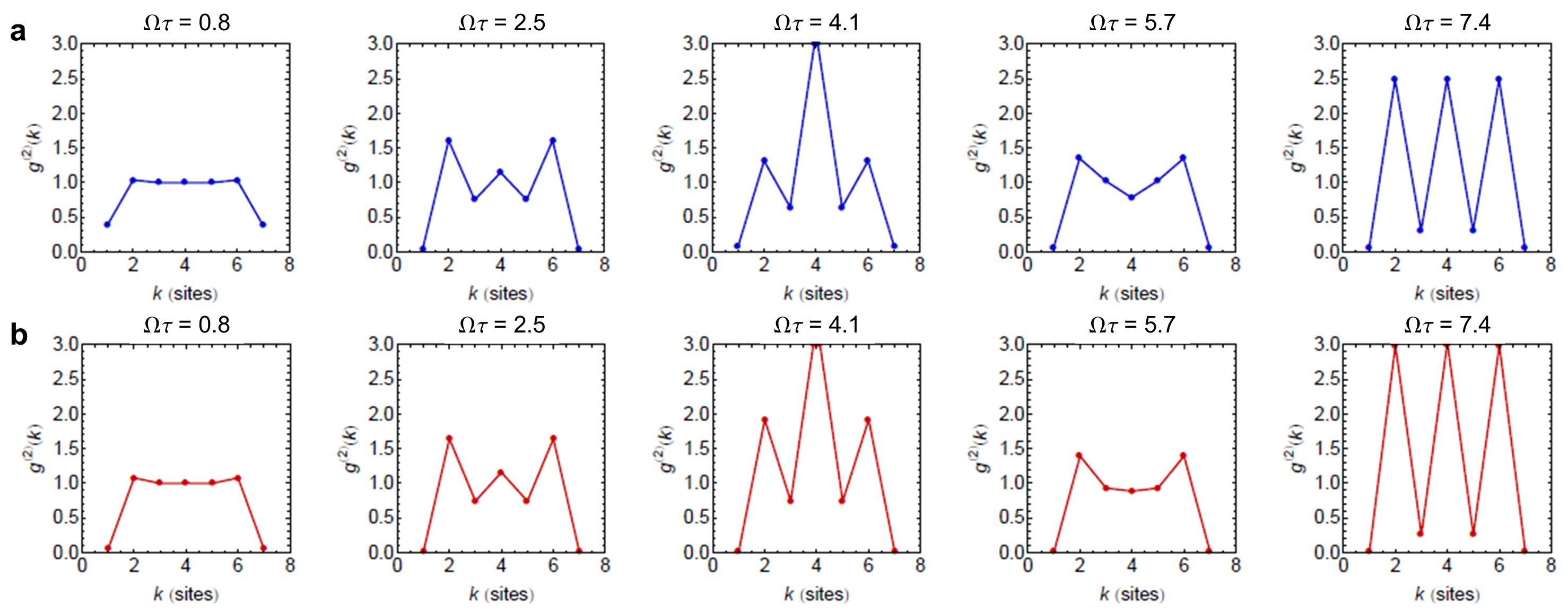}
\caption{\small {\bf Assessing the validity of the approximation of translational invariance in the 8-atom ring.} Calculated pair correlation function $g^{(2)}(k)$ as a function of the excitation time, for the 8-atom ring. {\bf a}: simulation using the experimentally relevant anisotropic interaction, which breaks translational invariance. {\bf b}: simulation with the same parameters as in {\bf a}, except that the angular dependence is neglected (we replace (\ref{eq:veff}) by its value for $\theta=0$), thus reestablishing translational invariance. One observes that the contrast in {\bf a} is reduced, as expected, but only in a marginal way.}
\label{fig:figs3}
\end{figure}

For the one-dimensional configurations of the main text (8-atom ring of Fig.3b and racetrack-shaped array of 30 traps of Fig.4a of main text) we plot the spatially averaged pair correlation function
\begin{equation}
g^{(2)}(k)=\frac{1}{N_{\rm t}}\sum_{i}\frac{\langle n_{i} n_{i+k}\rangle}{\langle n_{i} \rangle \langle n_{i+k} \rangle},
\label{eq:s:g2}
\end{equation}
where the subscripts label sites. For a system invariant by translation, all terms in the sum are identical, and the averaging over $i$ simply improves the signal to noise ratio. However, our systems are not translationally invariant, in particular because of the anisotropy of the interaction, and a natural question to address is whether the averaging reduces the contrast of the correlation functions. To answer this question, we have calculated the dynamics of the pair correlation function for the 8-atom ring, taking or not into account the anisotropy of the interaction (Extended Data Figure~\ref{fig:figs3}). One observes that the contrast reduction due to averaging is very small, thereby validating our choice to perform it for the data shown in the main text.

\subsection*{Effective loss mechanism arising from anisotropic interactions of D states} \label{sec:dephasing}

The agreement between our measurements and the results of the simulations is not perfect for the largest excitation times, in particular for some settings (e.g. for some configurations in the full blockade regime, for the 8-atom ring in the partial blockade regime, and for the $7\times7$ square array), where we observe a gradual increase in the number of measured Rydberg excitations. 

These effects could be qualitatively reproduced if the detection errors $\varepsilon$ would increase in time. However, the main reason for these losses is due to the fact that the microtraps are switched off during the excitation (to avoid inhomogeneous light-shifts), and as they are off \emph{for a fixed amount of time} (3~$\mu$s), independent of $\tau$, we do not, at first sight, expect $\varepsilon$ to increase in time. One could imagine however that the presence of the Rydberg excitation lasers may induce extra loss (due to off-resonant scattering for instance), and in this case one would end up having an $\varepsilon$ increasing with $\tau$. We have experimentally ruled out this possibility  by measuring the recapture probability when shining the Rydberg excitation lasers, detuned from the Rydberg line by $\sim 100~{\rm MHz}$, for the full 3~$\mu$s, without measuring any detrimental effect. 

A second possible reason would be the motion of the atoms. Due to their finite temperature, the atoms move during free flight with a velocity $v\sim 50\;{\rm nm/\mu s}$.  Now, strictly speaking, the terms corresponding to the laser coupling in Eqn. (1) of main text are not $\Omega \sigma_x^{i}$, but $\Omega {\rm e}^{i{\bs k}\cdot{\bs r}_i(t)} \sigma_+^{i}+{\rm h.c.}$, where ${\bs k}$ is the sum of the wavevectors of the excitation lasers at 795 and 475~nm, and ${\bs r_i}(t)$ the position of atom $i$. Thus, because of the motion, the phase factors of the couplings become time-dependent, which e.g. yields a dephasing of the spin wave corresponding to $\ket{W}$ states. However, a numerical simulation of this effect shows that the induced dephasing rates are negligible for our parameters. 

We thus believe that the cause for the observed extra losses lies in the interplay between the large number of interacting Zeeman sublevels when two atoms are excited to $nD_{3/2}$ states: for $\theta\neq0$ all 16 pair state Zeeman sublevels are coupled together by the van der Waals interaction. For a large number of atoms, this may lead to an effective loss rate from the targeted $\ket{r}$ states into a quasi-continuum comprising all other (weakly interacting) Zeeman states, and hence to a gradual increase of population of the Rydberg manifold. Qualitatively, this interpretation is corroborated by the fact that the observed increase in the number of observed excitations depends quite strongly on the array geometry: for instance, the data of the racetrack-shaped array, for which a majority of interacting atom pairs are almost aligned along the quantization axis $z$, are well reproduced by the simulations even at long times, unlike in the case of the 8-atom ring or the $7\times 7$ square array, for which many interacting pairs have their internuclear axes strongly inclined with respect to $z$.

Achieving a quantitative understanding of these observed imperfections, using approaches similar to the ones of refs.~{7,8,32}, is a challenging task. However, it is an important step in view of future applications of Rydberg blockade for quantum simulation, and will thus be the subject of future work.

\begin{enumerate}

\setcounter{enumi}{29}

\item 
Shen, C. \& Duan, L.-M.
\emph{Correcting detection errors in quantum state engineering through data processing}.
{New J. Phys. {\bf 14}, 053053 (2012)}.

\item 
Schau\ss, P., Zeiher, J., Fukuhara, T., Hild, S., Cheneau, M., Macr\`i, T., Pohl, T., Bloch, I. \& Gross, C.
\emph{Crystallization in Ising quantum magnets}.
{Science {\bf 347}, 1455 (2015)}.

\item 
Derevianko, A., Komar, P., Topcu, T.,  Kroeze, R.M. \& Lukin, M.D.
\emph{Effects of molecular resonances on Rydberg blockade}.
{arXiv:1508.02480}.

\end{enumerate}

\end{document}